\documentclass[aps,prl,twocolumn,floatfix,showpacs,preprintnumbers,amsmath,amssymb,nofootinbib]{revtex4-1}
\usepackage[T1]{fontenc}
\usepackage{graphicx}
\usepackage{bm}
\usepackage{color}
\usepackage{dcolumn}
\usepackage{amssymb}

\newcommand{\fm}{{\rm\,fm}}
\newcommand{\MeV}{{\rm\,MeV}}

\newcommand{\fmc}{{\rm\,fm/c}}

\newcommand{\I}{\mathrm{i}}
\renewcommand{\vec}[1]{\mbox{\boldmath $#1$}}
\newcommand{\vecs}[1]{\mbox{\boldmath {\scriptsize $#1$}}}

\begin{document}

\title{Time-dependent density functional theory  with twist-averaged boundary conditions}

\author{B.~Schuetrumpf}
\affiliation{FRIB Laboratory, Michigan State University, East
  Lansing, Michigan 48824, USA} 
  
  \author{W.~Nazarewicz} 
\affiliation{Department of Physics and
  Astronomy and FRIB Laboratory, Michigan State University, East
  Lansing, Michigan 48824, USA} 
\affiliation{Faculty
  of Physics, University of Warsaw, 02-093 Warsaw,
  Poland}
  
\author{P.-G.~Reinhard}
\affiliation{Institut f\"ur theoretische Physik, Universit\"at Erlangen, D-91054 Erlangen, Germany}

\date{\today}

\begin{abstract}
Time-dependent density functional theory is widely used to describe
excitations of many-fermion systems. In its many applications, 3D
coordinate-space representation is used, and infinite-domain
calculations are limited to a finite volume represented by a box.  For
finite quantum systems (atoms, molecules, nuclei), the commonly used
periodic or reflecting boundary conditions introduce spurious
quantization of the continuum states and artificial reflections from
boundary; hence, an incorrect treatment of evaporated
particles. These artifacts can be practically cured by introducing
absorbing boundary conditions (ABC) through an absorbing potential in
a certain boundary region sufficiently far from the described system.
But also the calculations of infinite matter (crystal electrons,
quantum fluids, neutron star crust) suffer artifacts from a finite computational box. In
this regime, twist-averaged boundary conditions (TABC) have been used
successfully to diminish the finite-volume effects.  In this work, we
extend TABC to time-dependent framework and apply it to resolve the
box artifacts for finite quantum systems using as test case small- and
large-amplitude nuclear vibrations. We demonstrate that by using such
a method, one can reduce finite volume effects drastically
without adding any additional parameters. While they are almost
equivalent in the linear regime, TABC and ABC differ in the nonlinear
regime in their treatment of evaporated particles.
\end{abstract}


\maketitle 

{\it Introduction} --- The time-dependent density functional theory
(TDDFT) for electronic systems had been developed as dynamical
extension of stationary DFT \cite{Dre90} in the early 1980ies
\cite{Run84} and has evolved in the meantime to a widely used,
efficient, and reliable tool to describe the dynamics of all sorts of
electronic systems, see
\cite{Marques06,Gro96}  for a review of the basics, and \cite{Rei03a,Furche2005,Mar04,Marques06}  for examples of applications.  A parallel development  took place in nuclear physics where TDDFT is known
under the notion of time-dependent Hartree-Fock (TDHF) scheme.  TDHF
as such was proposed as early as 1930 in \cite{Dirac}. Applications to
nuclei started in the mid 1970ies when appropriate computing
facilities became available \cite{Engel75,Bon76a,Cus76a}.  The
ever-improving computational capabilities had led to a revival of TDHF
without symmetry restrictions, applied to both finite systems
\cite{Nakatsukasa2005,Maruhn2005,Umar06a,Simenel12,Umar15} and
infinite matter under astrophysical conditions
\cite{Schuetrumpf2014,Schuetrumpf2015}.  The numerical tool of choice
for truly dynamical processes are coordinate- or momentum-space
representations of wavefunctions and fields and there exists a great
variety of published codes using these techniques for electronic
systems \cite{Cay06aR,Gia09aR} as well as for nuclear TDHF
\cite{Mar15a}.

A problem pertaining to  all numerical solutions of TDDFT
 is that one is bound to use finite basis sets. For example,
the most widely used scheme is based on the coordinate-space
representation of wave functions, densities, and potentials. The size of the box,
in which computations are carried out, is finite, and this implies that  either reflecting or periodic boundary
conditions (PBC) are imposed \cite{Mar15a}.  This leads to unphysical artifacts.  One
problem is that particles which are in principle emitted from the
system and thus traverse  the box boundaries are coming back to
the system area (reflected or reentering the simulation box from the
opposite side) and so perturb the dynamical evolution.  Moreover,   due to the presence of finite  box,  the continuum
states are artificially discretized, and this produces artifacts  at energies above continuum
threshold. Green's functions methods allow to cope with this problem in
the regime of linear response \cite{Shl75a,BertschBroglia94}. In case of grid
representations, one has to work on the boundary conditions.
Outgoing, or radiation,  boundary conditions which exactly connect the dynamics on
the grid to free flow in outer space are proposed as solution
\cite{Keller1989,Bou97,Man98a}, but they are very elaborate and hard to implement
in fully three dimensional grid representations. An efficient and
practical method are ABC, which were
introduced first in atomic calculations \cite{Kulander1987,Krause1992}
and are meanwhile also used in nuclear TDHF
\cite{Nakatsukasa2005,Reinhard2006,Pardi2013}. Although they can be implemented
technically with different algorithms, they amount in practice to
adding an imaginary potential in a certain boundary region. The
quality of the absorption depends on the profile of the imaginary
potential and its width \cite{Reinhard2006}. A good working compromise
has to be found in each application anew in order to suppress unwanted
remaining reflections as much as necessary.

Problems with finite simulation boxes appear also in calculations of
infinite matter. Periodic boundary conditions are appropriate in this
case, and yet, the wave functions are forced to be strictly periodic
which induces spurious quantization effects. This can be avoided by
TABC \cite{Gros1992,Gros1996,Lin},  often referred to as  `integration over boundary conditions'.  According
to the Floquet-Bloch theorem, a wave
function in a periodic potential is periodic up to a complex phase
shift (twist) when going from one cell to the next. Averaging over different
phase shifts very efficiently suppresses unwanted spurious
quantization effects \cite{Chiesa06,Matveeva12,Shulenburger13,Sorella15,Mostaani15}.  The benefits of TABC have also been
demonstrated in nuclear physics, including  time-independent simulations of infinite nucleonic
matter \cite{Carter,Chamel2007,Gandolfi09,Gulminelli,Hagen14,Schuetrumpf2015b} and  lattice QCD \cite{Bedaque05,Tiburzi05,Bernard11,Doring11,Briceno14,Briceno15}.  All these successful applications of TABC indicate that this method can help
 with the problem of the unphysically discretized continuum in
TDHF calculations of finite nuclei. This is the question, which  we aim
to investigate in this paper and we do that by comparing the
performance of TABC with that of ABC.

TABC is designed to suppress spurious finite-size quantization effects
and does that very well. It leaves, however, all particles in the
simulation box which means that the gas of emitted particles is still
around and may perturb system's dynamics. By employing
ABC, one can  avoid the gas because  the emitted
particles are removed efficiently. In the same way ABC help to reduce spurious finite-volume
quantization effects. However, imperfect absorption always leaves  some
quantum beating \cite{Reinhard2006}.  Moreover, ABC also absorb  the
outer tails of  bound-state wave functions; hence,  a faint
background of spurious particle emossion is produced   (to be avoided by sufficiently
large boxes).  As no practical prescription is perfect,  we have to balance advantages
and disadvantages of various ways of implementing boundary conditions.

{\it Boundary conditions} --- 
TABC are realized by implementing the Bloch boundary conditions
\begin{equation}\label{eq:bbc}
  \psi_{\alpha\vecs{\theta}}(\vec{r}+\vec{T}_i)
  =
  e^{\I\theta_i}\psi_{\alpha\vecs{\theta}}(\vec{r}),
\end{equation}
where $\vec{\theta}$ are three phases or twist angles, $\vec{T}_i$ ($i\in\{x,y,z\}$) is one of
the lattice vectors, $\psi_{\alpha\vecs{\theta}}(\vec{r})$ is the sigle-particle wave function characterized by the label
$\alpha$. When employing TABC, one
runs separately  DFT calculations with different twists  $\vec{\theta}$
and averages the results.  This can also be applied to
the TDDFT case. An observable to be evaluated is averaged according to
\begin{equation}
  \langle \hat{O}(t)\rangle
  = {1\over 8\pi^3}
  \iiint\limits_0^{\hspace{12pt}2\pi}  d^3\vec{\theta}\,
  \langle \Psi_{\vecs{\theta}}(t)| \hat{O}|
  \Psi_{\vecs{\theta}}(t)\rangle.
\label{eq:average}
\end{equation}
Considering the spatial symmetry of the problem, it is sufficient to average over $\theta_i$ 
between 0 and $\pi$ in all three directions for the isoscalar E2 mode and in the $x$ and 
$y$ directions for the isovector E1 mode.  The 3D integration over  $\vec{\theta}$  is carried 
out using an $n$-point Gauss-Lagrange quadrature between 0 and $\pi$ and 
$2n$-point between 0 and 2$\pi$. The total
number of TABC TDHF calculations to be performed is thus $n^3$ for the isoscalar E2 mode and $2n^3$ for the isovector E1 mode. 
 The Slater determinants $|\Psi_{\vecs{\theta}}(t)\rangle$
are obtained through independent TDHF calculations with the different
sets of twist angles $\vec{\theta}$.
  
ABC can be realized by either introducing an imaginary absorbing
potential in a boundary zone or by applying a mask function after each
TDHF step. Both methods are equivalent and can be mapped into each other
\cite{Reinhard2006}. Here we use  the mask function $f(r)$. One masking step
reads $\psi_\alpha
  \rightarrow
  \psi_\alpha f(r)$ with $f(r)=1$ for $r\leq L/2-l_\mathrm{abs}$;
$f(r)=\cos\left(\frac{\pi}{2}\frac{r-L/2+l_\mathrm{abs}}{l_\mathrm{abs}}\right)^p$ 
for $ L/2-l_\mathrm{abs}<r\leq L/2$; and $f(r)=0$ for 
$r> L/2$, where  $L$ is the cubic box length and 
$l_\mathrm{abs}$ is thickness of the absorbing sphere.  Optimal values
of $p$ depend on grid spacing and size of time
step \cite{Reinhard2006}. Here we use $p=0.0675$ throughout. We perform
the calculations in two different boxes: ($L=32$\,fm,  $l_\mathrm{abs}=6$\,fm) or 
($L=40$\,fm, $l_\mathrm{abs}=10$\,fm). Mind that the absorbing zone is applied at all
sides such that the active zone without absorption has in both
cases the same radius of $L-2 l_\mathrm{abs}=20$\,fm.

{\it Method} --- 
Our calculations are done using the  3D
Skyrme-TDHF solver \textsc{Sky3d} which is based on an equidistant, Cartesian 3D grid
\cite{Mar15a}.  We use it with a grid spacing of $\Delta x=1\fm$ and
time steps of $0.1\fmc$.  We use the Skyrme energy density functional 
SV-bas \cite{SVbas}.  The natural boundary conditions for the
plane-wave representation used are PBC. Note that
the long range Coulomb force is treated exactly (i.e., yielding
non-periodic $1/r$ asymptotics) using a Green's function formalism
\cite{Eas79}. Here  we have extended the code to accomodate ABC and TABC. Our benchmarking calculations are performed for
electric  dipole and quadrupole oscillations of
$^{16}$O.
The oscillations are triggered by an initial boost
  $\psi_{\alpha}(\vec{r}) \rightarrow \psi_{\alpha}(\vec{r})\,e^{-\I\eta F(\vecs{r})}$
where $F$ is the electric isovector dipole (E1) or isoscalar quadrupole (E2)
operator and $\eta$ is the excitation strength \cite{Rei07c}. This boost augments
the stationary ground-state wave functions  with a
velocity field which, in turn, drives dynamics. The observable we look
at is the emerging time evolution of the multipole moment
$\langle{F}\rangle$, from which we also produce the spectral
distribution of the multipole excitation strength, or power spectrum, by the windowed
Fourier transform of the time signal \cite{Cal97a}.

\begin{figure}[htb]
\includegraphics[width=\linewidth]{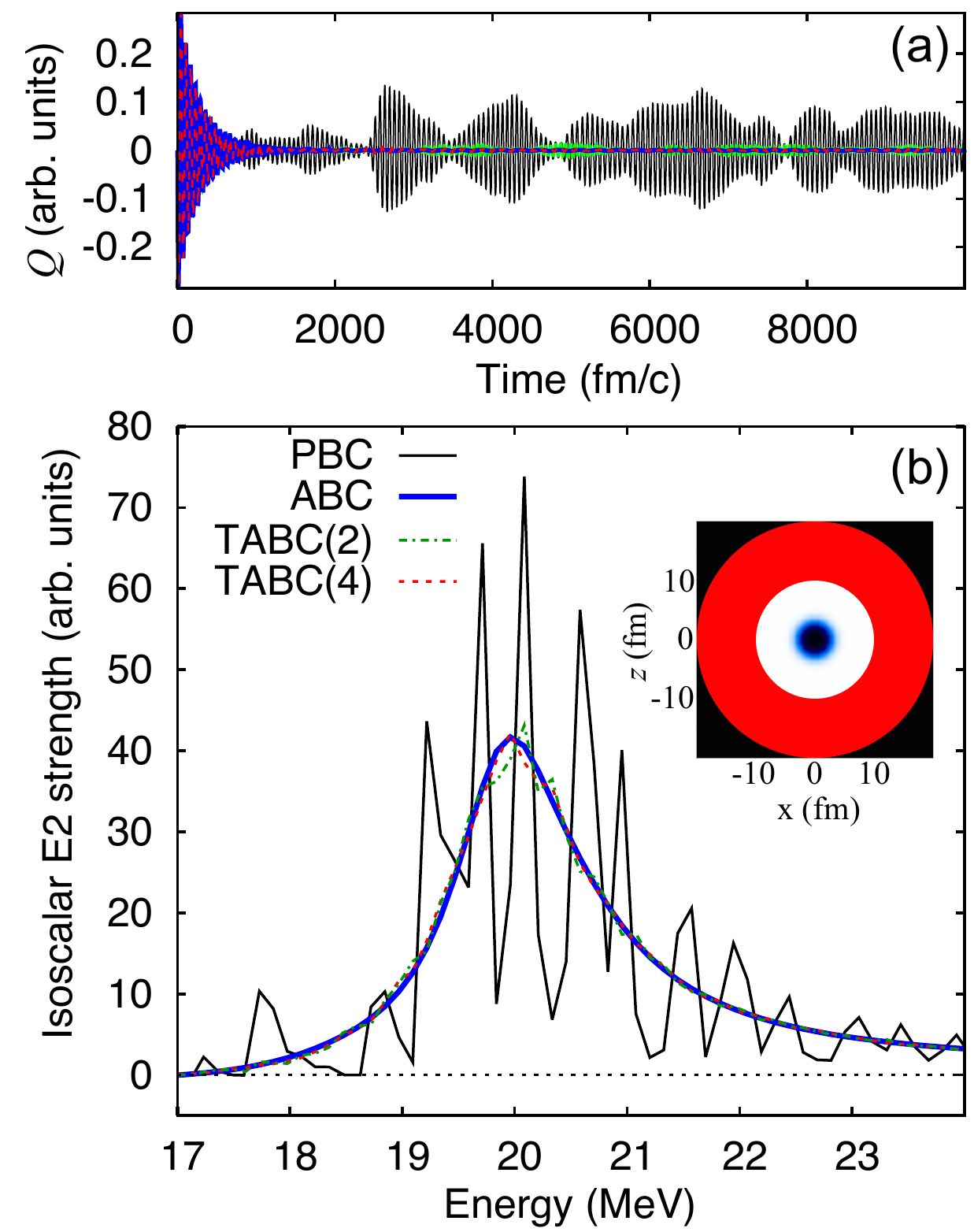}%
\caption{\label{fig:quad}(Color online) Quadrupole moment $Q(t)$ (a) and strength function (b) for the isoscalar E2 excitation of $^{16}$O  with $E^*=3\MeV$ and $L=40\fm$. The inset illustrates the geometry of the problem: the total density of $^{16}$O (center), the absorption zone (red), and the  region of zero density (black).}
\end{figure}

{\it Results} --- 
Figure~\ref{fig:quad} shows  the isoscalar E2 response
at the low excitation energy $E^*=3$\,MeV.  Compared are PBC, ABC and TABC($n$) results,  where $n$ denotes the number
of twists per direction. The
time signal in Fig.~\ref{fig:quad}(a) reveals that
PBC induce large-amplitude unphysical beating pattern after about
$1000\fmc$. The magnitude of these reverberations is as large as half of the maximum
amplitude at $t=0$. The use of ABC completely extinguishes them. While with
TABC(2) there are still some small spurious oscillations, with
TABC(4) the damping appears almost the same as with ABC. The corresponding strength functions are
displayed in Fig.~\ref{fig:quad}(b). The results obtained with PBC exhibit large fluctuations due to
the discretized continuum. With only $n=2$ points in each direction,  these oscillations are almost gone in TABC(2). The results in TABC(4) and ABC variants yield
 smooth quadrupole strength distributions and both
curves are practically the same.  Since TABC(4) represent a good compromise
between feasibility and accuracy, $n=4$ is therefore chosen for all
following calculations.

Although their results look very similar, the mechanism damping the signals
are much different in ABC and TABC variants. With ABC,  the erratic nucleon gas is removed
 from the box whenever it encounters the boundaries.
In TABC, the gas remains in the box as the particle number is strictly conserved by $\Psi_{\vecs{\theta}}(t)$ and every single run for given twist
shows qualitatively the same reverberations as PBC.  However, these
fluctuations enter  averaged quantities  (\ref{eq:average}) with different phases  and so average out.

\begin{figure}[tb]
\includegraphics[width=\linewidth]{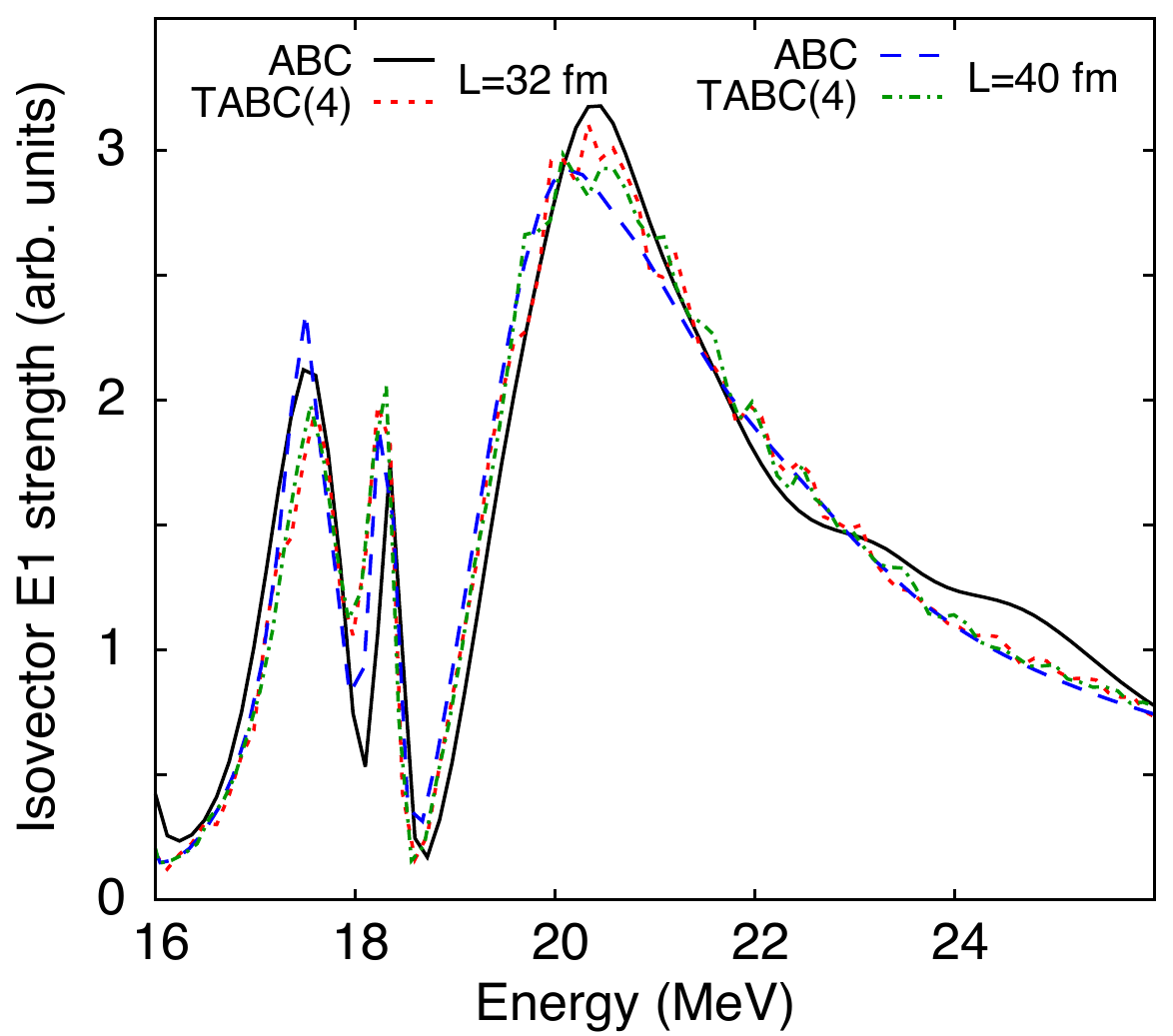}%
\caption{\label{fig:dip}(Color online) Isovector E1 strength  for $^{16}$O with $E^*=1\MeV$ and two box sizes: $L$=32 and 40\,fm.}
\end{figure}
We now move to the isovector E1 mode.  The corresponding strength function is shown in Fig.~\ref{fig:dip} for the low
excitation energy $E^*$. To study the dependence of results on box size, we compare
results obtained with  $L=32\fm$ and $L=40\fm$ by keeping the inner region of ABC (no masking function is applied) the same.
The PBC results (not shown) exhibit
the spurious finite-volume oscillations, which strongly depends on the box size.
As the excitation energy is small, there is a  small loss of about 0.1
nucleons in the ABC variant. Overall, ABC and TABC(4) calculations produce fairly similar strength functions for both box sizes. The enhanced shoulder around
$24\MeV$ for $L=32\fm$ in ABC is  a faint remainder of the artificial quantization of the
continuum \cite{Reinhard2006}. This feature is wiped out by the
improved absorption with $L=40\fm$. Another difference appears at the main peak at about 20.5
MeV where both TABC calculations agree aside from small remaining fluctuations 
which could be  further reduced by averaging over more twist points. The maximum for ABC and $L=40\fm$ appears at 
a lower energy and is lower as compared to $L=32\fm$.

\begin{figure}[tb]
\includegraphics[width=\linewidth]{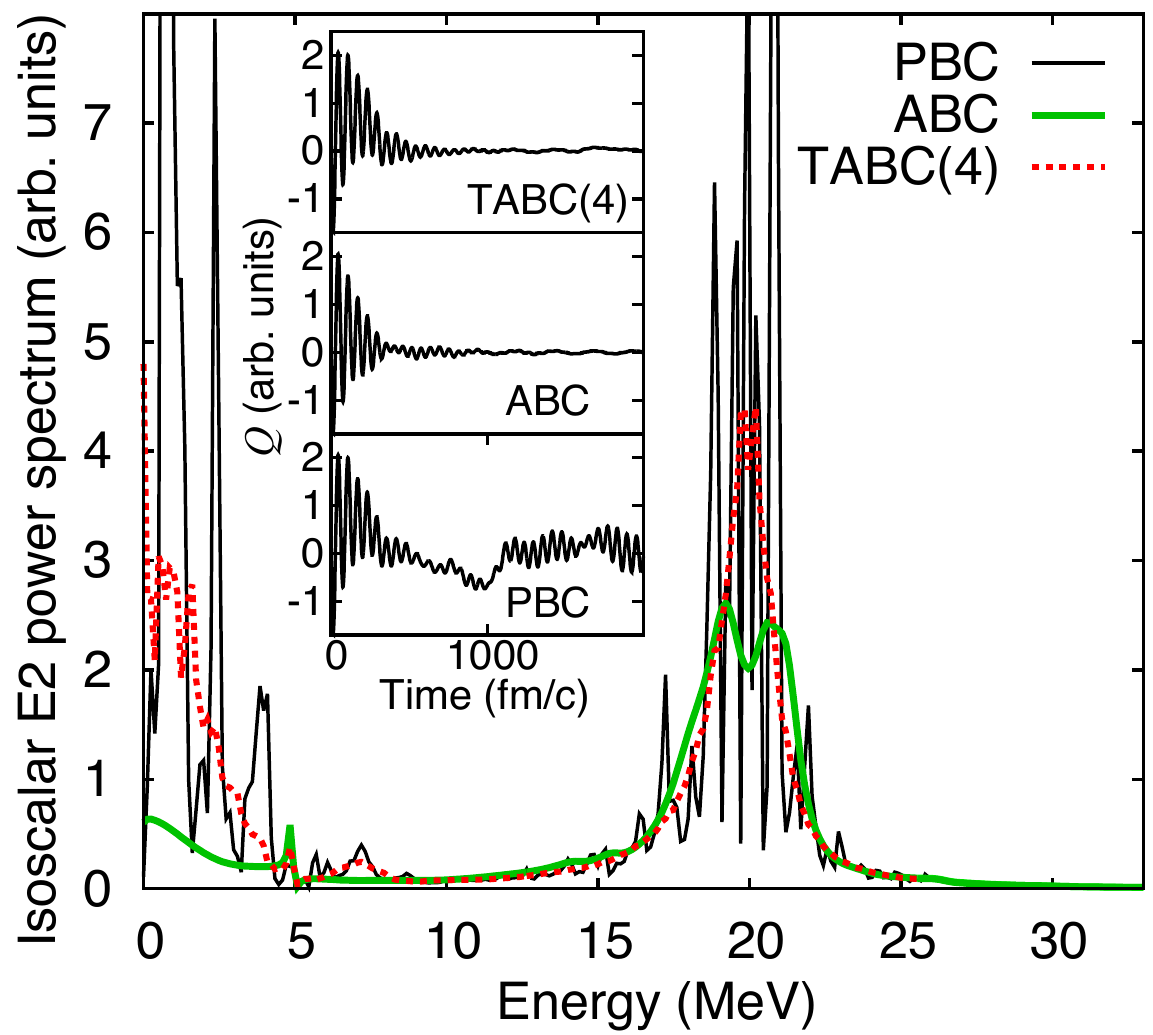}%
\caption{\label{fig:highquad}(Color online) 
Isoscalar E2 power spectrum for  $^{16}$O  at $E^*=20\MeV$ and
  $L=32\fm$. The inset shows the time evolution of the mass quadrupole moment $Q$. See \cite{SM} for animations.}
\end{figure}
We now proceed to higher excitation energies, $E^*\approx$20\,MeV. In this nonlinear regime
for $^{16}$O \cite{Rei07c},  we look at the power spectrum.
Here we take a strictly periodic dipole and quadrupole operators by replacing the coordinates with the periodic substitutes as in Ref.~\cite{Mar15a}.
Figure~\ref{fig:highquad} shows results for the isoscalar E2  excitation.
In the time signal, PBC shows the pronounced beat pattern stemming from low-frequency oscillations of  nucleonic gas moving within the full volume of the box. This effect is well visible in the E2 mode as the quadrupole moment, being quadratic in $x$, $y$, and $z$, is sensitive to the border areas of the box far from the oscillating nucleus. The clouds of oscillating nucleonic gas can be clearly seen in the animations included in Supplemental Material (SM) \cite{SM}.
Those spurious long-time
fluctuations are efficiently wiped out in both ABC and TABC after $t\approx 500$\,fm/c. 
At shorter times, say  the first 200 fm/c, we
see  some interesting differences in the time signal of TABC and ABC.
The initial TABC signal takes a bit longer to decay and the negative amplitudes are suppressed indicating that
the evaporated  particles are emitted predominantly  in the direction of positive quadrupole moment ($z$-direction). As seen in the animations in SM, the gas particles are absorbed efficiently in ABC, which results in a more symmetric time response.

\begin{figure}[tb]
\includegraphics[width=\linewidth]{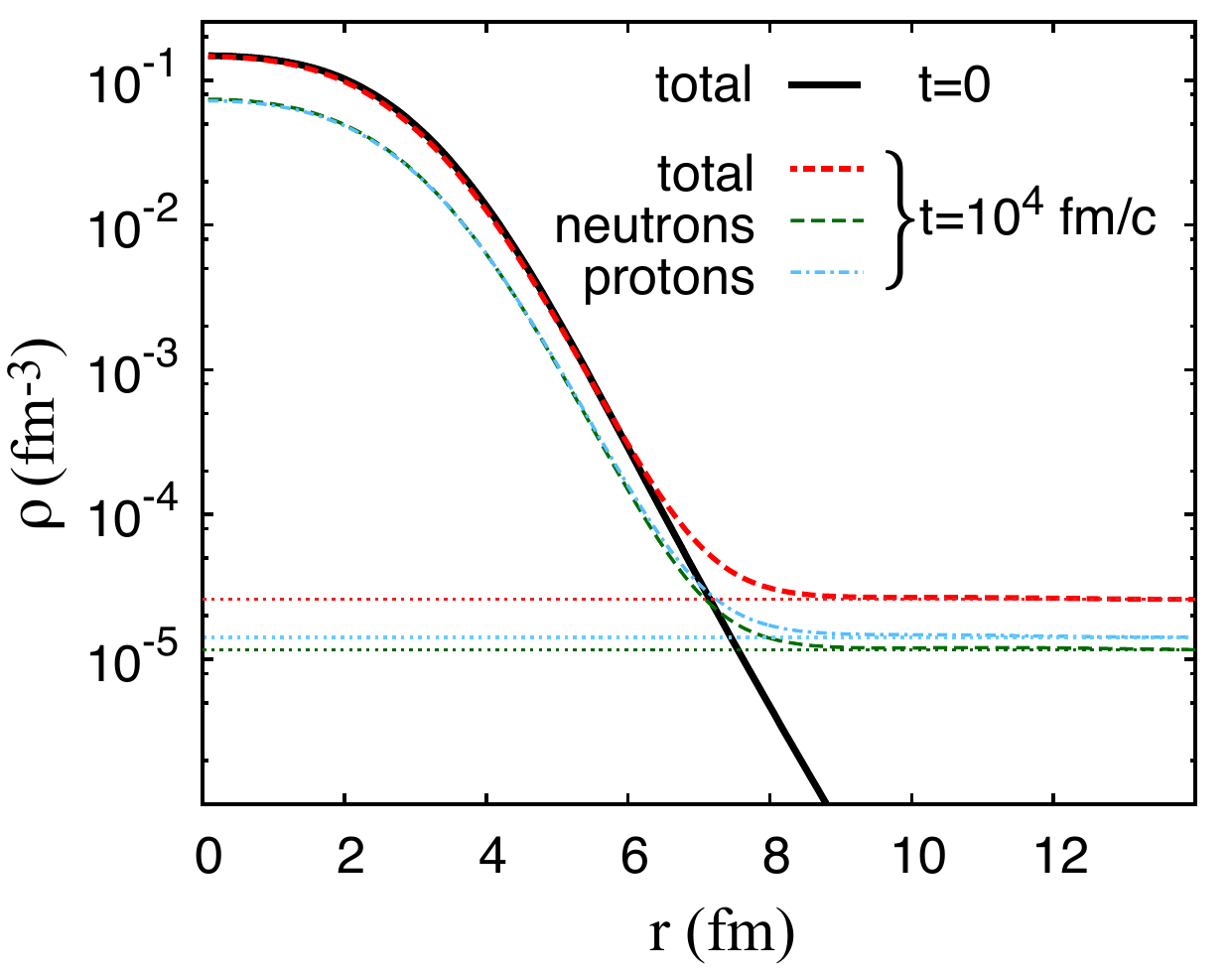}%
\caption{\label{fig:gas}(Color online) Angular-averaged density distribution corresponding to  the isoscalar E2 mode at  $E^*=20\MeV$ at two snaphots: $t=0$ and $t=10^4$\,fm/c. The gas densities  are indicated by  dotted lines: $\rho_{\rm n,gas}=1.15\cdot10^{-5}$fm$^{-3}$ and
$\rho_{\rm p,gas}=1.42\cdot10^{-5}$fm$^{-3}$.
}
\end{figure}

To understand the effect of the background gas in TABS, Fig.~\ref{fig:gas} shows angular-averaged density distributions associated with E2 vibrations of Fig.~\ref{fig:highquad}. The averaging was done by means of Gaussians centered at mesh points of the original Cartesian 3D grid. At times longer than $500\fmc$, the resulting  nucleonic gas is uniformly distributed within the volume of the box. 
The magnitude of the gas density carries information about the effective temperature of the system and the particle emission rate \cite{Kerman81,Bonche84,Bonche85,Pei14}.
Integrating the gas density results in  0.38 neutrons and 0.47 protons, and this nicely agrees with the number of absorbed particles  in the ABC variant.

The power spectrum for PBC shows  large
fluctuations in the resonance region around $20\MeV$ and huge
spikes at energies $<4\MeV$. Those peaks are removed by ABC and TABC
 and replaced by smooth low-energy bumps associated with nucleonic gas.
In the TABC variant,  the number of particles is strictly conserved and the gas is kept in the box. This results in a pronounced low-frequency effect.  In  ABC, absorbing potential removes most of the gas efficiently leaving only the
unavoidable effect from a loosely bound nucleon halo.  There is also a small difference in the
resonance spectra around $20\MeV$ where a dip appears with ABC. This can be attributed
to an effect from the nucleon gas in TABC, which disturbs the dynamics in the resonance region.

 \begin{figure}[tb]
\includegraphics[width=\linewidth]{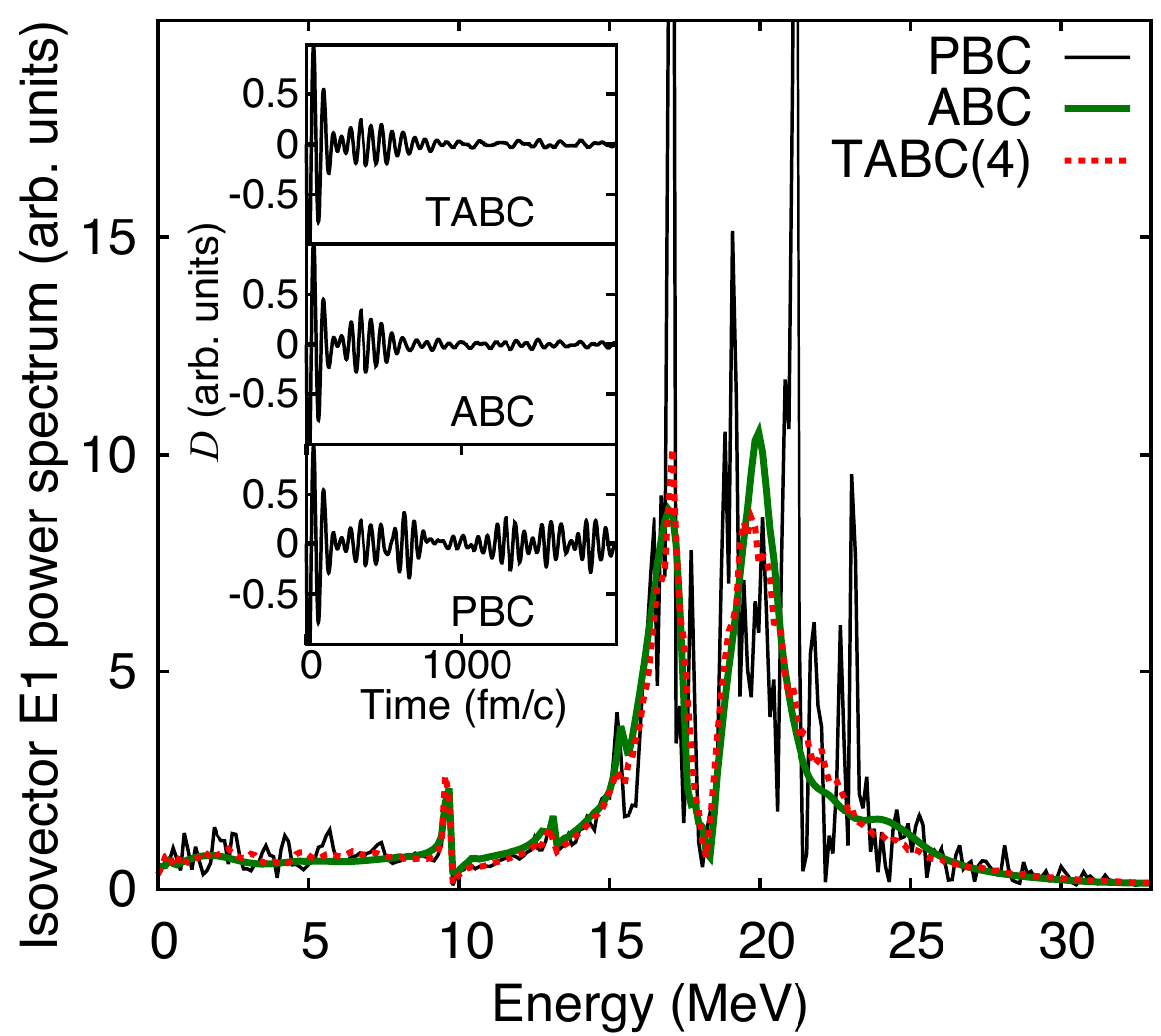}%
\caption{\label{fig:highdip}(Color online) Isovector E1 power spectrum for  $^{16}$O  at $E^*=22\MeV$ and
  $L=32\fm$. The inset shows the time evolution of the dipole moment $D$. See \cite{SM} for animations.}
\end{figure}
The agreement between ABC and TABC is much closer for the E1 mode shown in 
Fig.~\ref{fig:highdip}.  We see again that ABC as well
as TABC remove the reverberation in the time signal and the spurious
fluctuations in the spectrum.  Both  approaches reproduce nicely
the strong enhancement of the low-energy dipole peak at about $10\MeV$
as well as the flat profile down to $E=0$.

{\it Conclusions} ---
We demonstrated that TABC can be implemented into TDDFT framework and tested it for nuclear vibrations. Adding no additional parameters, the new approach  removes spurious finite-volume effects as efficiently as the previously used method based on ABC.    With only two  twist phases per direction one obtains a reasonable  reduction of  spurious fluctuations;  four twists per direction offer a good compromise between feasibility and quality. Since TABC calculations for different twists can be performed independently, the method is easily adapted to parallel computing. 

For low-energy excitations corresponding to  the linear regime, TABC give very similar results as ABC.  In the non-linear regime, ABC absorb noticeable parts of wave
 functions which for TABC remain in the box as floating nucleon
 gas. Nonetheless, we see a good agreement. Both methods suppress
 efficiently the box artifacts and provide very similar spectra,
 except for some difference in the quadrupole response at very low
 energies where TABC shows a sizeable bump associated with
 the slow long-range fluctuations of the nucleon gas. ABC is more
 efficient in suppressing this artifact.

In future applications, the new TDDFT+TABC method will be applied to  excitations of heavy, superfluid nuclei. Furthermore, we intend to apply  TABC to infinite systems such as nuclear pasta oscillations in the neutron star crust. In this case, the nucleonic gas represents physical reality, and the low-frequency bump associated with the motion of the cloud within the box is likely to impact the transport properties of the crust.

\begin{acknowledgements}
This material is based upon work supported by the U.S. Department of Energy, Office of Science under Award Numbers DOE-DE-NA0002847 (the Stewardship Science Academic Alliances program) and DE-SC0008511 (NUCLEI SciDAC-3 collaboration). This work used computational resources of  the Institute for Cyber-Enabled Research at  Michigan State University.
\end{acknowledgements}

\bibliography{TABC,pgr}

\end{document}